\documentclass[journal]{IEEEtran}
\usepackage{enumitem}
\usepackage{graphicx}
\usepackage{cite}
\usepackage{array}
\usepackage{makecell}
\usepackage{multirow}
\usepackage{hhline}
\usepackage{float}
\usepackage{url}
\ifCLASSINFOpdf
\else
\fi
\begin{document}
\title{Real-time data compression for data acquisition systems applied to the ITER Radial Neutron Camera}

\author{B.~Santos, N.~Cruz, A.~Fernandes, P.F.~Carvalho, J.~Sousa, B.~Gon\c{c}alves, M.~Riva, F.~Pollastrone, C.~Centioli, D.~Marocco, B.~Esposito, C.M.B.A.~Correia and R.C.~Pereira

\thanks{Manuscript submitted in June 15, 2018. 
The work leading to this publication has been funded partially by Fusion for Energy under the Contract F4E-FPA-327. IST activities also received financial support from “Funda\c{c}\~{a}o para a Ci\^{e}ncia e Tecnologia” through project UID/FIS/50010/2013. This publication reflects the views only of the author, and Fusion for Energy cannot be held responsible for any use which may be made of the information contained therein.}
	
\thanks{B. Santos, A. Fernandes, N. Cruz, P.F. Carvalho, J. Sousa, B. Gonçalves, R.C. Pereira, are with Instituto de Plasmas e Fus\~{a}o Nuclear, Instituto Superior T\'{e}cnico, Universidade de Lisboa, 1049-001 Lisboa, Portugal (e-mail:bsantos@ipfn.tecnico.ulisboa.pt).}

\thanks{M. Riva, F. Pollastrone, C. Centioli, D. Marocco, B. Esposito are with ENEA C. R. Frascati, Dipartimento FSN, via E. Fermi 45, 00044 Frascati (Roma), Italy.}
\thanks{C.M.B.A. Correia is with LIBPhys-UC, Department of Physics, University of Coimbra, P-3004 516 Coimbra, Portugal.}
}


\markboth{504 Real-time data compression for data acquisition systems applied to the ITER Radial Neutron Camera}%
{Shell \MakeLowercase{\textit{et al.}}: Bare Demo of IEEEtran.cls for IEEE Journals}



\maketitle

\begin{abstract}
To achieve the aim of the ITER Radial Neutron Camera Diagnostic, the data acquisition prototype must be compliant with a sustained 2 MHz peak event for each channel with 128 samples of 16 bits per event. The data is acquired and processed using an IPFN FPGA Mezzanine Card (FMC-AD2-1600) with 2 digitizer channels of 12-bit resolution and a sampling rate up to 1.6 GSamples/s mounted in a PCIe evaluation board from Xilinx (KC705) installed in the host PC.

\par The acquired data in the event-based data-path is streamed to the host through the PCIe x8 Direct Memory Access (DMA) with a maximum data throughput per channel $\approx$0.5 GB/s of raw data (event base), $\approx$1 GB/s per digitizer and up to 1.6 GB/s in continuous mode.

\par The prototype architecture comprises an host PC with two KC705 modules and four channels, producing up to 2 GB/s in event mode and up to 3.2 GB/s in continuous mode. To reduce the produced data throughput from host to ITER archiving system, the real-time data compression was evaluated using the LZ4 lossless compression algorithm, which provides compression speed up to 400 MB/s per core.

\par This paper presents the architecture, implementation and test of the parallel real-time data compression system running in multiple isolated cores. The average space-saving and the performance results for long term acquisitions up to 30 minutes, using different data block size and different number of CPUs, is also presented.
\end{abstract}


% \fi
\IEEEpeerreviewmaketitle

\section{Introduction} \label{introduction}

\IEEEPARstart{T}{his} Radial Neutron Camera (RNC) is a key ITER diagnostic aiming at the real-time measurement of the neutron emissivity to characterize the neutron emission that will be produced by the ITER tokamak \cite{marocco_2012,marocco_2016,mriva_2018,rcpereira_2015, rcpereira_2017, ncruz_2017}. To achieve the aim of the RNC diagnostic, the data acquisition prototype must be compliant with a sustained 2 MHz peak event for each channel with 128 samples of 16 bits. The data is acquired and processed using IPFN FPGA Mezzanine Cards (FMC-AD2-1600) with 2 digitizer channels of 12-bit resolution sampling up to 1.6 GSamples/s. These in-house developed cards are mounted in PCIe evaluation boards from Xilinx (KC705) \cite{rcpereira_2017}.

\par The prototype architecture comprises one host PC with two installed KC705 modules and four channels, producing up to 2 GB/s in event mode and up to 3.2 GB/s in continuous mode \cite{rcpereira_2017}.

\par The LZ4 is a lossless compression algorithm, providing compression speed at 400 MB/s per core, scalable with multi-core CPUs. This algorithm appears as the fastest compression algorithm with a relevant compression ratio comparing to other dictionary encoding and entropy encoding algorithms \cite{collet_lz4,kuhn_2016}.

\par During the RNC diagnostic prototype phase, the LZ4 was chosen to evaluate the feasibility of the real-time data compression implementation in the host PC to reduce the produced data throughput to ITER archiving system \cite{ncruz_2017}.

\par
LZ4 is also suitable for implementation in FPGAs \cite{bartik_2015,liu_2018} or in the Graphics Processor Units (GPUs) such as other lossless algorithms \cite{patel_2012, ozsoy_2011,ozsoy_2012,sitaridi_2014}, which can be a valuable feature for future developments.

\par This paper presents the implemented solution and the achieved results, which contribute to the RNC diagnostic specification. A brief overview of the system and software architecture is provided in Section \ref{systemarchitecture}. The preliminary results that contributes to the design of the implemented architecture are presented in Section \ref{preliminarytests}. Section \ref{results} presents the tests and results with the developed solution and selected compression algorithm. The paper ends with Section \ref{conclusions} devoted to the conclusions and future work remarks.

\section{System Architecture} \label{systemarchitecture}

\begin{figure}[!htb]
	\centering
	\includegraphics[width=1\linewidth]{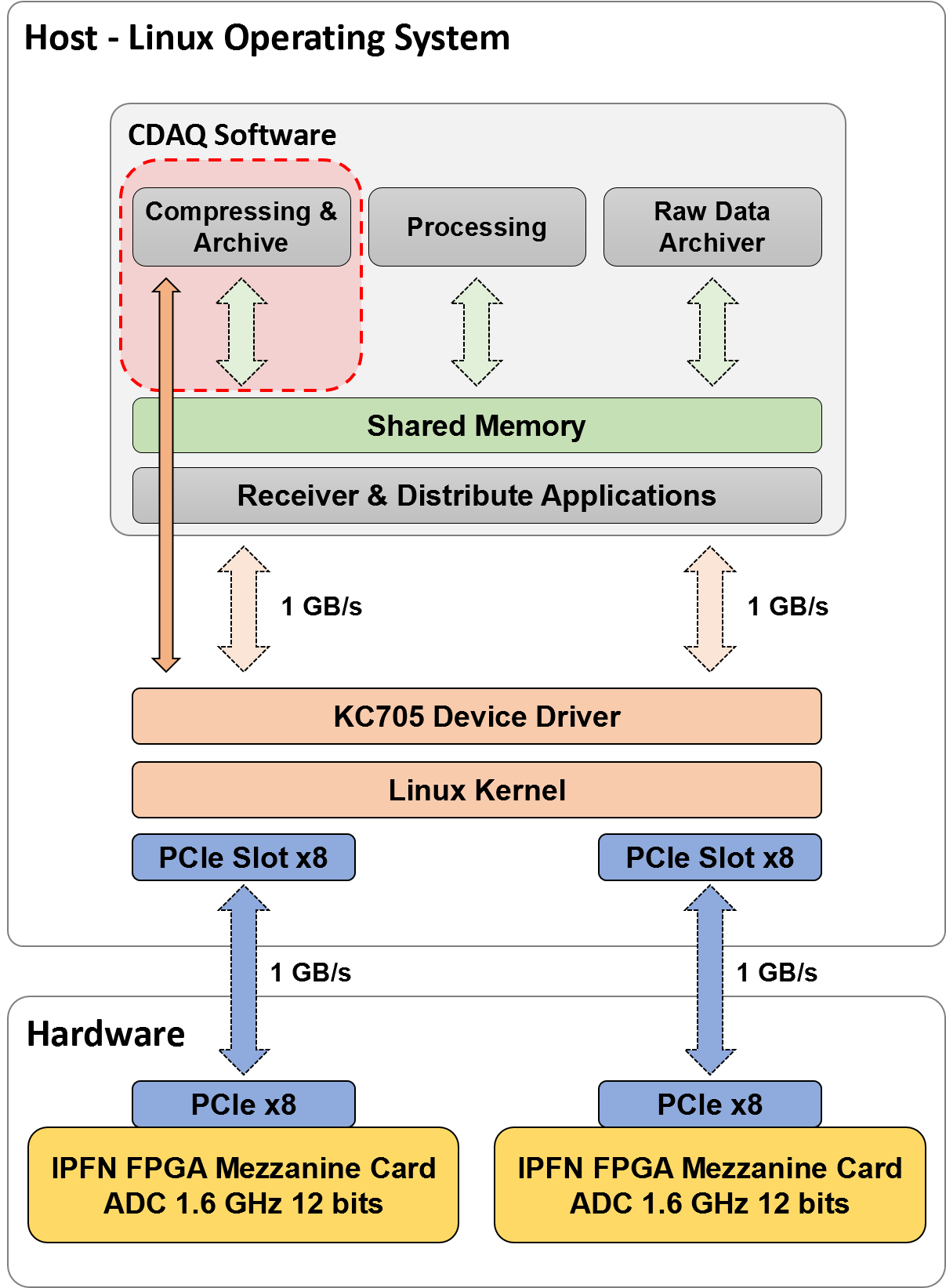}
	\caption{System Architecture}
	\label{fig:sysarch}
\end{figure}

Fig. \ref{fig:sysarch} depicts the overall system architecture \cite{ncruz_2018}, highlighting the context compression software path. The system was designed to support two IPFN FPGA Mezzanine Cards installed in the PCIe evaluation boards from Xilinx (KC705) and connected to the host through the PCIe x8 slots. The data production using a down-sampled configuration to 400 MSamples/s is up to 1 GB/s per board in event mode and can be increased to 1.6 GB/s per board in continuous mode \cite{afernandes_2018}, providing higher data throughput to stress tests.

\par The host computer hardware specification includes:
\begin{itemize}[leftmargin=+.5in]
	\item Motherboard: ASUS Rampage V Extreme with 4xPCIe 3.0/2.0 x16 slots
	\item CPU: Intel\textsuperscript{\textregistered} Core\textsuperscript{TM} i7-5930K@3.50 GHz supporting Intel\textsuperscript{\textregistered} Hyper-Threading Technology (6 cores, 12 threads)
	\item 64 GB of RAM and 256 GB SSD.
\end{itemize}
\par The Scientific Linux 7 is running as Operating System with kernel 3.10-rt and LZ4 version 1.7.5. 

\par Interfacing between the hardware and high-level applications is installed the Linux device driver, supporting data transfers up to 1.6 GB/s per board. The Control and Data Acquisition (CDAQ) software includes a shared memory layer to distribute the acquired real-time data across the client applications when several clients need to use the data at same time.

\par At the application level, software modules were developed for:
\begin{itemize}[leftmargin=+.5in]
	\item Real-time data compression to reduce the data size.
	\item Data pulse processing for energy and particle discrimination.
	\item Real-time raw archiving for test purposes with low data rate acquisitions.
\end{itemize}

\par In the presented tests, the compression application was directly connected to the device driver for evaluating its performance limit.

\par The introduction of the shared memory layer should not affect performance since the approach to read the data from the shared memory must be, in the worst case, as fast as the direct reads from the device drive.

\subsection{Software Architecture}
\begin{figure}[!htb]
	\centering
	\includegraphics[width=1\linewidth]{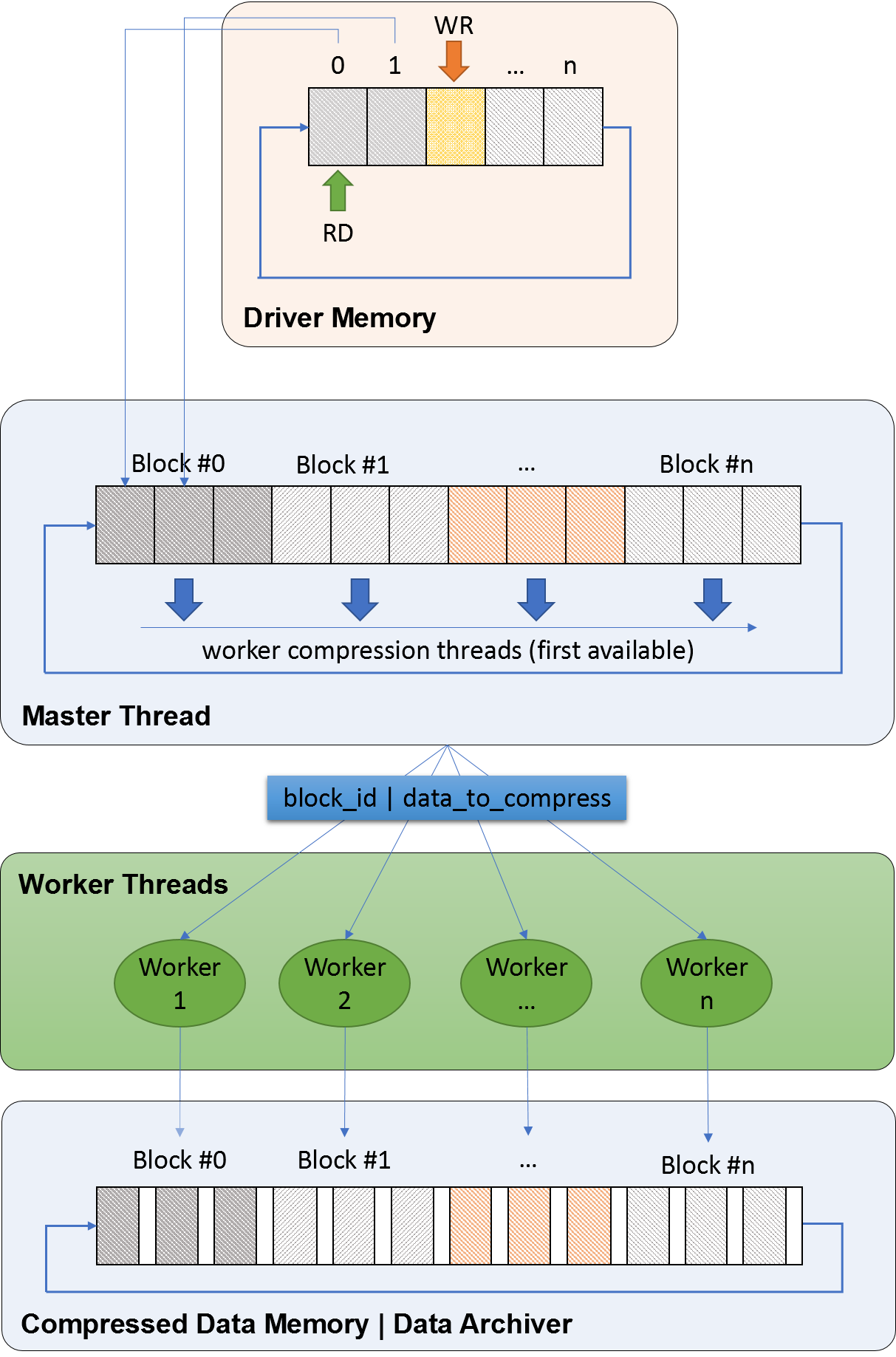}
	\caption{Software Architecture}
	\label{fig:softarch}
\end{figure}

Fig. \ref{fig:softarch} presents the compression application software architecture and its interface to the device driver. The compression application is based on the task farm algorithm approach with a master thread that launches a poll of threads with a configured number of worker threads.

\par The device driver implements a kernel thread with an internal circular buffer to store in real-time the data transferred from the hardware, until it is read from the consumer applications. There are two implemented pointers for the circular buffer (read pointer and write pointer) to control the read operations from the consumer applications and checks data loss between the device driver and the consumer applications. Additionally, the device driver implements algorithms to check the data loss between the hardware and device driver.

\par The master thread reads the available data from device driver in real-time, packs it into data blocks with configurable data size and distributes it across the configured number of worker threads. Each data block is tagged with an id to be used by the worker threads to store the compressed data in the correct position of a shared buffer between worker threads. The work balance algorithm distributes the next block to be compressed to the first available core in the pool but other options can be used.

\par The worker threads implement the compression algorithm and are responsible for the parallel compression in real-time. In the present tests, the worker threads store the compressed data in a memory buffer but can deliver it through the network directly to the data archiver.

\par The device driver, master thread and worker threads run in isolated cores (detached from the kernel scheduling, preventing its usage by the operating system), taking advantage of the CPU affinity feature, which is the ability to direct a specific task, or process, to use a specified core.

\section{Preliminary Tests} \label{preliminarytests}
To identify the achieved compression speed, compression ratio and space-saving with different configurations of the LZ4 algorithm several tests are done using the LZ4 default and LZ4 HC. In the LZ4 default an acceleration option can be configured to get a better compression speed compromising the compression ratio. On the LZ4 HC, a high compression derivative of LZ4, the compression level can be configured to improve the compression ratio compromising the compression speed.

\par The input data for these tests was collected with real radation sources in Frascatti Neutron Generator (FNG) during the tests in January 2018 and from a waveform generator simulating a gamma ray type signal as input to compare different signal types. The acquisition was configured with a pulse width of 128 samples, producing 256 MB/s throughput.

\par Table \ref{Table_LZ4_default} presents the LZ4 default compression tests comparing the different sources and acceleration factors. The results suggest that the accelerations with even number (2, 4, 6, 8, 10 and 12) have a better relation between compress speed and compress ratio. However, the acceleration with factor 1 has the better compression ratio.

%\IEEEpubidadjcol

\begin{table}[!htb]

	\caption{LZ4 Default Tests}
	\label{Table_LZ4_default}	
	\centering	
	%\begin{tabular}{c | C{0.8cm} | C{0.5cm} | C{0.6cm} | C{0.8cm} | C{0.5cm} | C{0.6cm}|}
	\def\arraystretch{1.40}
	\setlength\tabcolsep{0pt}
	\begin{tabular}{|c|ccc|ccc|}	
		\hline
		\hline		
		\multirow{3}{ 1.80cm}{\centering Acceleration}  										 &
		 \multicolumn{3}{ c|}{\centering FNG}                                            & \multicolumn{3}{ c|}{\centering PULSE SIGNAL}                                   \\ 
		\cline{2-7}
		\multicolumn{1}{|m{1.80cm}|}{}         &
		\multicolumn{1}{ m{1.20cm}}{\centering Speed (MB/s)}         & 
		\multicolumn{1}{ m{1.00cm}}{\centering Ratio}                & 
		\multicolumn{1}{ m{1.10cm}|}{\centering Space Saving}         & 
		\multicolumn{1}{ m{1.20cm}}{\centering Speed (MB/s)}         & 
		\multicolumn{1}{ m{1.00cm}}{\centering Ratio}                & 
		\multicolumn{1}{ m{1.10cm}|}{\centering Space Saving}         \\ 
		\hline	
		
		1                    & 337.25               & 1.38                 & 27.63\%              & 317.39               & 1.44                 & 30.75\%              \\
		2                    & 419.15               & 1.34                 & 25.22\%              & 400.51               & 1.40                 & 28.31\%              \\
		3                    & 409.85               & 1.32                 & 24.43\%              & 385.96               & 1.35                 & 26.04\%              \\
		4                    & 475.87               & 1.29                 & 22.15\%              & 470.52               & 1.31                 & 23.44\%              \\
		5                    & 448.09               & 1.28                 & 21.82\%              & 429.99               & 1.29                 & 22.72\%              \\
		6                    & 523.47               & 1.25                 & 19.97\%              & 524.37               & 1.26                 & 20.53\%              \\
		7                    & 492.34               & 1.25                 & 19.84\%              & 477.26               & 1.25                 & 19.96\%              \\
		8                    & 575.08               & 1.21                 & 17.41\%              & 603.26               & 1.20                 & 16.75\%              \\
		9                    & 539.40               & 1.21                 & 17.53\%              & 533.68               & 1.20                 & 16.97\%              \\
		10                   & 600.72               & 1.19                 & 16.05\%              & 662.61               & 1.17                 & 14.56\%              \\
		11                   & 566.24               & 1.19                 & 16.13\%              & 580.35               & 1.18                 & 15.33\%              \\
		12                   & 659.45               & 1.15                 & 13.32\%              & 727.40               & 1.15                 & 12.79\%              \\ 
		\hline
		\hline
	\end{tabular}
\end{table}

\par Table \ref{Table_LZ4_HC} presents the LZ4 High Compression (HC) tests comparing the different sources and compression levels. Using the 256 MB/s of data throughput with up to 8 CPU cores running in parallel, only the first 3 compression levels can be used without missing data. The results present a better space saving than in the LZ4 default but compromising significantly the compression speed.

\begin{table}[!htb]
	
	\caption{LZ4 High Compression Tests}
	\label{Table_LZ4_HC}	
	\centering	
	%\begin{tabular}{c | C{0.8cm} | C{0.5cm} | C{0.6cm} | C{0.8cm} | C{0.5cm} | C{0.6cm}|}
	\def\arraystretch{1.40}
	\setlength\tabcolsep{0pt}
	\begin{tabular}{|c|ccc|ccc|}			
		\hline
		\hline		
		\multirow{3}{ 1.80cm}{\centering Compression Level}  										 &
		\multicolumn{3}{ c|}{\centering FNG}                                            & \multicolumn{3}{ c|}{\centering PULSE SIGNAL}                                   \\ 
		\cline{2-7}
		\multicolumn{1}{|m{1.80cm}|}{}         &
		\multicolumn{1}{ m{1.20cm} }{\centering Speed (MB/s)}         & 
		\multicolumn{1}{ m{1.00cm} }{\centering Ratio}                & 
		\multicolumn{1}{ m{1.10cm}|}{\centering Space Saving}         & 
		\multicolumn{1}{ m{1.20cm} }{\centering Speed (MB/s)}         & 
		\multicolumn{1}{ m{1.00cm} }{\centering Ratio}                & 
		\multicolumn{1}{ m{1.10cm}|}{\centering Space Saving}         \\ 
		\hline		
		1                             & 89.39                 & 1.54                  & 35.22\%               & 101.87                & 1.56                  & 36.06\%               \\ 
		2                             & 73.07                 & 1.62                  & 38.20\%               & 83.31                 & 1.63                  & 38.56\%               \\ 
		3                             & 55.28                 & 1.69                  & 40.67\%               & 58.79                 & 1.71                  & 41.40\%               \\ 
		4*                            & 40.97                 & 1.74                  & 42.62\%               & 32.23                 & 1.79                  & 44.15\%               \\ 
		5*                            & 30.76                 & 1.78                  & 43.95\%               & 20.70                 & 1.86                  & 46.24\%               \\ 
		6*                            & 23.25                 & 1.81                  & 44.82\%               & 13.00                 & 1.91                  & 47.55\%               \\ 
		7*                            & 17.84                 & 1.83                  & 45.33\%               & 9.96                  & 1.93                  & 48.06\%               \\ 
		8*                            & 14.58                 & 1.84                  & 45.54\%               & 9.78                  & 1.92                  & 47.96\%               \\ 
		9*                            & 13.75                 & 1.84                  & 45.59\%               & 9.35                  & 1.93                  & 48.07\%               \\ 
		10*                           & 12.47                 & 1.84                  & 45.63\%               & 9.31                  & 1.93                  & 48.17\%               \\ 
		11*                           & 9.66                  & 1.86                  & 46.28\%               & 7.58                  & 1.96                  & 48.88\%               \\ 
		12*                           & 9.60                  & 1.86                  & 46.28\%               & 7.52                  & 1.96                  & 48.93\%               \\ 
		\hline
		\hline
	\end{tabular}
\end{table}

\par Table \ref{Table_LZ4_Cores} presents the theoretically number of needed cores to compress 1 GB/s of data throughput in real-time with different LZ4 default acceleration factors and different LZ4 HC compression levels.

\begin{table}[!htb]
	\caption{Number of Cores for each Compression Configuration}
	\label{Table_LZ4_Cores}	
	\centering	
	\def\arraystretch{1.40}
	\setlength\tabcolsep{0pt}	
	\begin{tabular}{|c|c|c|c|c|}
		\hhline{==~==}
		\multicolumn{2}{|c|}{\centering LZ4 Default} & \multicolumn{1}{l|}{} & \multicolumn{2}{c|}{\centering LZ4 HC} \\ \cline{1-2} \cline{4-5} \cline{1-2} \cline{4-5} 
		\multicolumn{1}{|m{1.80cm}|}{\centering Acceleration} &
		\multicolumn{1}{ m{1.50cm}|}{\centering Cores}        &   
		\multicolumn{1} {m{1.00cm} }{\centering ~}                    & 
		\multicolumn{1}{|m{1.80cm}|}{\centering Compression Level}  & 
		\multicolumn{1}{ m{1.50cm}|}{\centering Cores}  \\ 
		\cline{1-2} \cline{4-5} 
		1                   & 4           &                       & 1                  & 13     \\  
		2                   & 3           &                       & 2                  & 16     \\  
		3                   & 3           &                       & 3                  & 22     \\  
		4                   & 3           &                       & 4*                 & 33     \\  
		5                   & 3           &                       & 5*                 & 49     \\  
		6                   & 2           &                       & 6*                 & 69     \\  
		7                   & 3           &                       & 7*                 & 84     \\ 
		8                   & 2           &                       & 8*                 & 107    \\  
		9                   & 3           &                       & 9*                 & 113    \\ 
		10                  & 2           &                       & 10*                & 115    \\ 
		11                  & 2           &                       & 11*                & 134    \\  
		12                  & 2           &                       & 12*                & 136    \\ 
		\hhline{==~==}
	\end{tabular}
\end{table}

\par Using the LZ4 default algorithm there is no difference in the needed cores between the acceleration levels 2 and 5, but the space saving reduces $\approx$5\%. Using the compression level 1, the space saving is increased $\approx$2\% but one more core is need.

\par The LZ4 high compression variant usage is not possible because it needs at minimum 13 available cores (10 more than LZ4 default) to improve the space saving in $\approx$6\%.

\par The tests also confirm compressing speed and ratio similarities between the acquired signals in the real environment and signals from the waveform generator.

\section{Tests and Results} \label{results}
The tests were based in a pulse type signal from a waveform generator simulating a gamma-ray distribution. The acquisition tests have different pulse width configurations to produce distinct acquisition data rates up to 1.5 GB/s. Each acquisition test has 60 minutes, in agreement with the ITER long pulse acquisitions. All tests have 10 MB of data block size to compress, except the tests to verify the impact of the usage of other block sizes.

\par To compress data with maximum compression ratio available, the LZ4 default algorithm with acceleration factor 1 was selected.

\subsection{Number of Cores}
Table \ref{Table_Data_Loss} presents the relation between the data loss and number of cores for different data acquisition rates.

\begin{table}[H]
	\centering
	\caption{Data Loss}
	\label{Table_Data_Loss}
	\def\arraystretch{1.40}
	\setlength\tabcolsep{0pt}
	\begin{tabular}{c|cccccc}
		\hline
		\hline
		\multirow{2}{ 1.60cm}{\centering Data Rate (MB/s)}  &
		\multicolumn{6}{c}{\centering Number of Cores}                     \\
		\multicolumn{1}{ m{1.60cm}|}{\centering ~} &
		\multicolumn{1}{ m{1.15cm} }{\centering 1} &
		\multicolumn{1}{ m{1.15cm} }{\centering 2} &   
		\multicolumn{1}{ m{1.15cm} }{\centering 3} & 
		\multicolumn{1}{ m{1.15cm} }{\centering 4} & 
		\multicolumn{1}{ m{1.15cm} }{\centering 5} &
		\multicolumn{1}{ m{1.15cm} }{\centering 6}   \\ 
		\hline
		128                               & 0.00\%  & -       & -       & -       & -      & -      \\
		256                               & 0.00\%  & -       & -       & -       & -      & -      \\
		512                               & 40.68\% & 0.00\%  & -       & -       & -      & -      \\
		768                               & 61.06\% & 22.33\% & 0.00\%  & -       & -      & -      \\
		1024                              & 69.38\% & 38.83\% & 8.80\%  & 0.00\%  & -      & -      \\
		1536                              & 79.58\% & 59.27\% & 39.09\% & 19.12\% & 4.32\% & 0.00\\
		\hline
		\hline
	\end{tabular}
\end{table}

\par Based on the results, the minimum number of needed cores per data acquisition rate with no data loss are:
\begin{itemize}[leftmargin=+.5in]
	\item 1 core to compress until 256 MB/s
	\item 2 cores for 512 MB/s
	\item 3 cores for 768 MB/s
	\item 4 cores for 1 GB/s
	\item 6 cores for 1.5 GB/s.
\end{itemize}

\par These results are inside the range of the preliminary results for acceleration level 1, which have a compression speed $\approx$300 MB/s per core. 

\subsection{CPU Usage}
The used CPU supports Intel\textsuperscript{\textregistered} Hyper-Threading Technology, providing 12 logical cores for the operating system, based on their 6 physical cores. This architecture can result in slightly differences to tests with 12 dedicated cores that can produce small improvements.

\par Table \ref{Table_Work_Load} presents average core usage of the dedicated cores to the master and worker threads for each acquisition data rate.

\begin{table}[!htb]
	\centering
	\caption{CPU Usage}
	\label{Table_Work_Load}
	\def\arraystretch{1.20}
	\setlength\tabcolsep{0pt}
	\begin{tabular}{ccc|cc}
		\hline
		\hline
		\multicolumn{1}{ m{1.60cm} }{\centering Pulse Width (Number of Channels)} &
		\multicolumn{1}{ m{1.40cm} }{\centering Data Rate (MB/s)} &
		\multicolumn{1}{ m{1.40cm}|}{\centering Number of Cores} &   
		\multicolumn{1}{ m{1.40cm} }{\centering Worker Threads} & 
		\multicolumn{1}{ m{1.40cm} }{\centering Master Thread}   \\ 
		\hline
		32 (1 Ch) & 128   & 1     & 37.00\% & 0.54\% \\
		64 (1 Ch) & 256   & 1     & 81.00\% & 0.59\% \\
		128 (1 Ch) & 512   & 2     & 84.00\% & 0.57\% \\
		256 (1 Ch) & 768   & 3     & 86.00\% & 0.56\% \\
		128 (2 Ch) & 1024  & 4     & 82.00\% & 2.75\% \\
		256 (2 Ch) & 1536  & 6     & 88.00\% & 25.64\% \\
		\hline
		\hline
	\end{tabular}
\end{table}

\subsection{Compression Statistics}
Table \ref{Table_Comp_Stats} summarizes the compression results of the tests with different pulse widths.

\begin{table}[!htb]
	\centering
	\caption{Compression Statistics}
	\label{Table_Comp_Stats}
	\def\arraystretch{1.30}
	\setlength\tabcolsep{0pt}
	\begin{tabular}{cc|ccccc}
	    \hline
		\hline
		\multirow{2}{ 1.80cm }{\centering Pulse Width (Number of Channels)} &
		\multirow{2}{ 1.30cm }{\centering Data Rate (MB/s)} & 
		\multirow{2}{ 1.10cm }{\centering Ratio} & 
		\multirow{2}{ 1.20cm }{\centering Space Saving} & 
		\multicolumn{3}{m{ 3.00cm }}{\centering Compression Speed (MB/s)} \\
		&                                   &                        &                               & 
		\multicolumn{1}{ m{1.10cm} }{\centering Avg} &
		\multicolumn{1}{ m{1.10cm} }{\centering Min} &
		\multicolumn{1}{ m{1.10cm} }{\centering Max}   \\
		\hline
		32 (1 Ch)                                         & 128                               & 1.33                   & 24.91\%                       & 353.73        & 342.20         & 355.63       \\
		64 (1 Ch)                                         & 256                               & 1.43                   & 29.97\%                       & 317.43        & 307.80         & 319.10        \\
		128 (1 Ch)                                        & 512                               & 1.55                   & 35.61\%                       & 302.76        & 294.87        & 304.23       \\
		256 (1 Ch)                                        & 768                               & 1.63                   & 38.47\%                       & 297.74        & 280.02        & 299.04       \\
		32 (2 Ch)                                         & 256                               & 1.33                   & 25.07\%                       & 352.35        & 340.90         & 353.41       \\
		64 (2 Ch)                                         & 512                               & 1.44                   & 30.54\%                       & 321.14        & 312.86        & 322.40        \\
		128 (2 Ch)                                        & 1024                              & 1.54                   & 35.17\%                       & 310.64        & 212.45        & 313.25       \\
		256 (2 Ch)                                        & 1536                              & 1.60                    & 37.36\%                       & 284.58       & 196.20        & 313.33   \\
		\hline
		\hline   
	\end{tabular}
\end{table}

\subsection{Block Sizes}
Table \ref{Table_Block_Sizes} presents the test of different data block sizes with same input signal and configurations (1 channel with pulse width 128 and acquisition data rate of 512 MB/s).
\begin{table}[!htb]
	\centering
	\caption{Block Sizes Comparison}
	\label{Table_Block_Sizes}
	\def\arraystretch{1.40}
	\setlength\tabcolsep{0pt}
	\begin{tabular}{c|ccccc}
		\hline
		\hline
		\multirow{2}{ 1.50cm }{\centering Block Size (MB)} & 
		\multirow{2}{ 1.10cm }{\centering Ratio} & 
		\multirow{2}{ 1.30cm }{\centering Space Saving} & 
		\multicolumn{3}{m{ 3.00cm }}{\centering Compression Speed (MB/s)} \\
		&                        &                               & 
		\multicolumn{1}{ m{1.10cm} }{\centering Avg} &
		\multicolumn{1}{ m{1.10cm} }{\centering Min} &
		\multicolumn{1}{ m{1.10cm} }{\centering Max}           \\
		\hline
		1                                        & 1.52        & 34.14\%        & 296.86                 & 269.57                        & 298.93\\
		10                                       & 1.55        & 35.61\%        & 302.76                 & 294.87                        & 304.23\\
		100                                      & 1.54        & 35.06\%        & 305.43                 & 304.82                        & 306.06\\
		200                                      & 1.52        & 34.29\%        & 306.36                 & 305.34                        & 306.77\\
		\hline
		\hline  
	\end{tabular}
\end{table}

\par The results suggest that data block size did not improve the space saving, however the standard deviation of the compression speed during the pulse is reduced.

\subsection{CPU and Memory Usage}
\begin{figure}[!htb]
	\centering
	\includegraphics[width=1\linewidth]{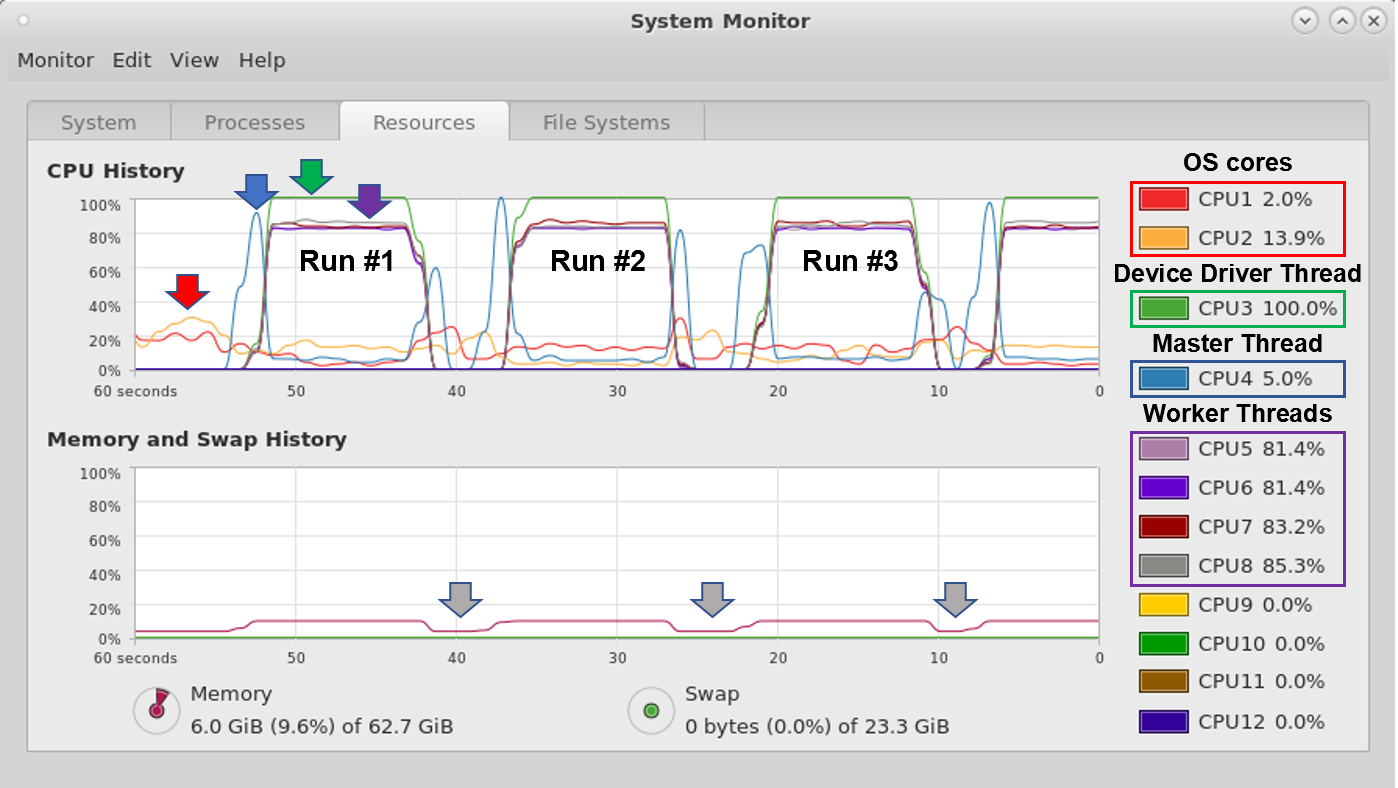}
	\caption{CPU and memory statistics with one board}
	\label{fig:cpumem1}
\end{figure}

Fig. \ref{fig:cpumem1} presents the CPU and memory usage during 3 acquisitions from one board during 10 seconds with 1024 MB/s of acquisition data rate. There are two dedicated cores for the operating system, one isolated core for the device driver thread, one isolated core for the compression master thread and four isolated cores for the compression worker threads. The allocated memory is around 6 GB in run-time to the data buffers.
\par
\begin{figure}[!htb]
	\centering
	\includegraphics[width=1\linewidth]{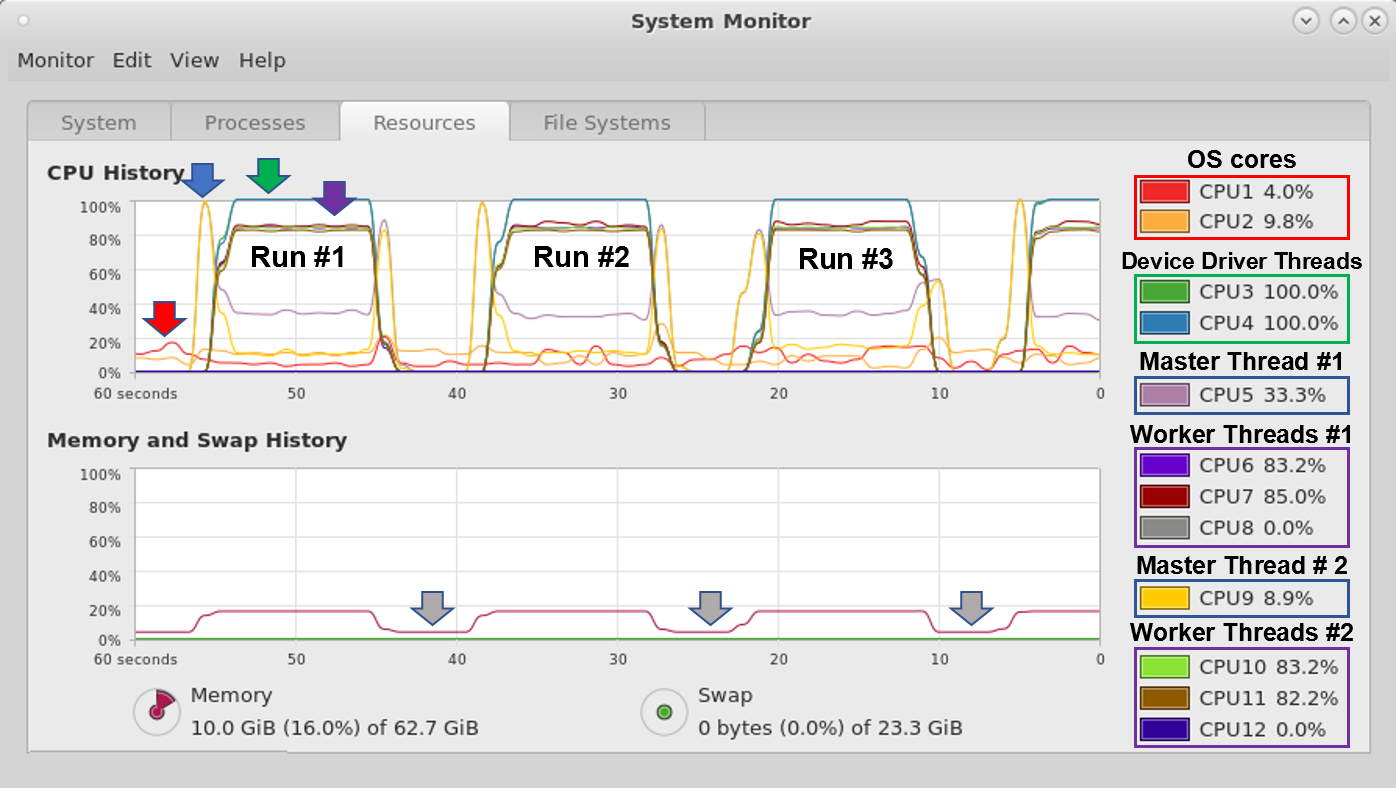}
	\caption{CPU and memory statistics with two boards}
	\label{fig:cpumem2}
\end{figure}

 Fig. \ref{fig:cpumem2} present the CPU and memory usage during 3 acquisitions from two boards simultaneously during 10 seconds with 512 MB/s of data acquisition rate per board. There are two dedicated cores for the operating system and each board uses five isolated cores (one for the device driver thread, one for the compression master thread and three for the worker threads). The allocated memory is around 10 GB in run-time to the data buffers.

\subsection{Relation Between Space Saving and Pulse Width}
\begin{figure}[H]
	\centering
	\includegraphics[width=1\linewidth]{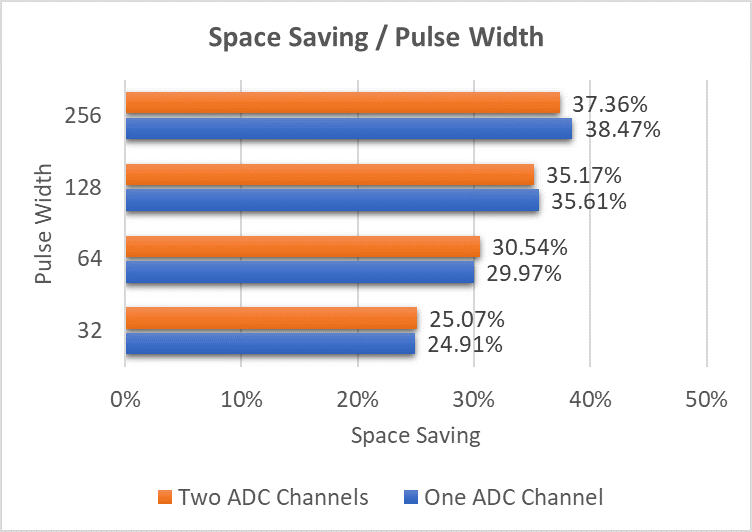}
	\caption{Average Space Saving per Pulse Width}
	\label{fig:spacesaving}
\end{figure}

Fig. \ref{fig:spacesaving} depicts the relation between space saving and pulse width for signals using one and two ADCs.
\par Independently of the ADCs number, a relation between space saving and pulse width can be identified. Using a greater pulse width, the relative space saving increases which can be related with the type of acquired data.

\section{Conclusions and Future Work} \label{conclusions}
This contribution evaluates the feasibility of data compression implementation in the host PC and contributes to the RNC diagnostic specification.

\par The presented architecture is scalable and adjustable. The number of worker threads can be configured to comply with different algorithms and data throughput.

\par The stress tests show a stable solution during 60-minute acquisitions with data acquisition rates up to 1.5 GB/s, using a maximum of 6 worker threads in parallel.

\par The system was also tested in Fedora Linux 27, kernel 4.16, the community version of Red Hat Linux that supports ITER CODAC system. The results were similar, which validates the developed architecture for future kernel version of Red Hat Linux.

\par Based on the presented tests, to compress 1 GB/s from one board in real-time, a minimum of 5 cores are needed (1 master and 4 worker threads). Using two boards simultaneously to acquire 1GB/s in each, the system will need 14 cores (2 cores for operating system, 2 cores for the device driver, 2 cores for master thread and 8 cores for the worker threads). There are several commercial CPUs that support this architecture.

\par The preliminary tests with two boards simultaneously showed a possible performance decreasing. Acquiring 512 MB/s the system needs 3 cores, instead of 2 with a single board. This can be related with the usage of Intel\textsuperscript{\textregistered} Hyper-Threading Technology instead of dedicated cores but intensive tests with two hardware modules acquiring simultaneously are scheduled to a future task.

\par With the tested signals, the maximum achieved space saving with the LZ4 algorithm was between 25\% and 40\%. Changes on signal configuration can influence the compress ratio.

\par In the future, the data compression can be implemented in the GPU or FPGA to compare the results with the host PC. There are some possible advantages to be tested but for the FPGA implementation, a new data path to the host is needed, once the processing algorithms need decompressed data in real-time. This increases the data transfer throughput between FPGA and host PC, which can be a demanding task for the Linux device driver.

\section*{Acknowledgment}
This manuscript is in memory of Professor Carlos Correia who is no longer among us.

\ifCLASSOPTIONcaptionsoff
  \newpage
\fi



% \bibliographystyle{IEEEtran}

\end{document}